\documentstyle[12pt]{article}
\begin{document}
\baselineskip 18pt
\def\today{\ifcase\month\or
 January\or February\or March\or April\or May\or June\or
 July\or August\or September\or October\or November\or December\fi
 \space\number\day, \number\year}
\def\thebibliography#1{\section*{References\markboth
 {References}{References}}\list
 {[\arabic{enumi}]}{\settowidth\labelwidth{[#1]}
 \leftmargin\labelwidth
 \advance\leftmargin\labelsep
 \usecounter{enumi}}
 \def\newblock{\hskip .11em plus .33em minus .07em}
 \sloppy
 \sfcode`\.=1000\relax}
\let\endthebibliography=\endlist
\def\lsim{\ ^<\llap{$_\sim$}\ }
\def\gsim{\ ^>\llap{$_\sim$}\ }
\def\r2{\sqrt 2}
\def\beq{\begin{equation}}
\def\eeq{\end{equation}}
\def\beqn{\begin{eqnarray}}
\def\eeqn{\end{eqnarray}}
\def\rmuu{\gamma^{\mu}}
\def\rmud{\gamma_{\mu}}
\def\PL{{1-\gamma_5\over 2}}
\def\PR{{1+\gamma_5\over 2}}
\def\sinW2{\sin^2\theta_W}
\def\AEM{\alpha_{EM}}
\def\mul{M_{\tilde{u} L}^2}
\def\mur{M_{\tilde{u} R}^2}
\def\mdl{M_{\tilde{d} L}^2}
\def\mdr{M_{\tilde{d} R}^2}
\def\mz2{M_{z}^2}
\def\c2b{\cos 2\beta}
\def\au{A_u}
\def\ad{A_d}
\def\cob{\cot \beta}
\def\v#1{v_#1}
\def\tb{\tan\beta}
\def\epem{$e^+e^-$}
\def\KK{$K^0$-$\bar{K^0}$}
\def\wi{\omega_i}
\def\xj{\chi_j}
\def\Wmu{W_\mu}
\def\Wnu{W_\nu}
\def\m#1{{\tilde m}_#1}
\def\mH{m_H}
\def\mw#1{{\tilde m}_{\omega #1}}
\def\mx#1{{\tilde m}_{\chi^{0}_#1}}
\def\mc#1{{\tilde m}_{\chi^{+}_#1}}
\def\mwi{{\tilde m}_{\omega i}}
\def\mxi{{\tilde m}_{\chi^{0}_i}}
\def\mci{{\tilde m}_{\chi^{+}_i}}
\def\mz{M_z}
\def\sw{\sin\theta_W}
\def\cw{\cos\theta_W}
\def\cb{\cos\beta}
\def\sb{\sin\beta}
\def\rwi{r_{\omega i}}
\def\rxj{r_{\chi j}}
\def\rfp{r_f'}
\def\Kik{K_{ik}}
\def\Fq2{F_{2}(q^2)}
\begin{titlepage}

\  \
\vskip 0.5 true cm 
\begin{center}
{\large {\bf The Neutron and the Electron  }}  \\
{\large {\bf Electric Dipole Moment  in N=1 Supergravity Unification}}
\vskip 0.5 true cm
\vspace{2cm}
\renewcommand{\thefootnote}
{\fnsymbol{footnote}}
 Tarek Ibrahim and Pran Nath
\\
\vskip 0.5 true cm 
\it Department of Physics, Northeastern University  \\
\it Boston, MA 02115, USA  \\
\end{center}

\vskip 4.0 true cm

\centerline{\bf Abstract}
\medskip
An analysis of the neutron EDM and of the electron EDM in minimal N=1
supergravity unification with two CP violating phases is given. For the
neutron the analysis includes the complete one loop gluino, chargino, and
neutralino exchange diagrams for the electric dipole and the chromoelectric
dipole operators, and also the contribution of the purely gluonic dimension
six operator. It is shown that there
exist significant regions in the six dimensional parameter space of the
model where 
cancellations between the gluino and the chargino exchanges reduce the 
electric and the chromoelectric contributions, and further cancellations
among the  electric, the chromoelectric, and the purely gluonic parts
lead to a dramatic lowering of the neutron EDM sometimes below the electron
EDM value. This phenomenon gives a new mechanism, i.e., that of internal
cancellations,  for the suppression of the neutron EDM in supersymmetric 
theories. 
The cancellation mechanism can significantly reduce 
the severe fine tuning problem associated  with CP violating phases in 
SUSY and SUGRA  unified models.

\end{titlepage}

\newpage 
\section{Introduction}

Supersymmetric models with softly  broken supersymmetry introduce
new sources of CP violation which contribute to the neutron and the 
electron electric dipole moment(EDM). 
It is known that the exchange of SUSY particles 
close to their current experimental lower limits and   
CP violating phases of normal size, i.e. O(1),
 will lead to the neutron EDM already in  excess
of the current experimental bound of $1.1\times 10^{-25}e$cm~\cite{altra}.
Two possibilities to resolve this problem have been commonly discussed 
in the literature. The first 
is that the phases are not O(1) but rather much smaller, i.e., 
$O(10^{-2}-10^{-3})$~\cite{ellis,dugan,bern}. 
However, a small phase constitutes
 a fine tuning unless it arises naturally, e.g. as a loop correction.
 The second 
possibility is that the phases are O(1), but the supersymmetric 
spectrum which contributes to the EDMs is heavy~\cite{na,kizu}, 
i.e. in the several TeV region and perhaps 
out of reach of even the LHC. In this paper we
 discuss a third possibility, i.e., that of internal cancellations 
among the different components of the neutron EDM.
 We shall show that such cancellations
 can dramatically reduce the neutron EDM without either excessive
  finetuning of  the
 phases  or pushing the SUSY spectrum in several TeV
 mass range. One then finds that the neutron and the electron
 EDMs can satisfy the current experimental bounds\cite{altra,commins}
  with phases not unduly
 small and a SUSY spectrum which is not unduly heavy at least for low 
 values of tan$\beta$.

	Although there are many analyses of the EDMs in supersymmetric
theories most of these~\cite{ellis,dugan,na,kizu,garis} are without radiative
breaking of the electro-weak symmetry and sometimes neglecting
the chargino contribution to the neutron EDM~\cite{ellis,garis}.
	  For the neutron EDM there are 
	two operators other than the electric dipole moment operator,
       which can contribute to the neutron EDM.
	One of these is the color dipole operator and the other is the
	dimension six purely gluonic operator
	considered by Weinberg\cite{wein2}. 
	 With the
	exception of the work of ref.\cite{arno2} most of the previous
analyses~\cite{ellis,dugan,kizu,garis,falk} do  not take into
	account the contribution  of the color and of
	the purely gluonic operators with the presumption that their
	relative contributions  to the neutron EDM is
	small. However,  it was pointed out in ref.\cite{us} that 
	 the contributions  of the color and of the purely gluonic operators
	 can be comparable to the contribution of the 
	 electric dipole  operator in a 
	significant 
region of the minimal supergravity parameter space~\cite{us}. 
Currently there is some confusion in the literature regarding the sign of the
gluino exchange term\cite{kizu,garis}.
Further, there is no analysis  aside from  that of ref.\cite{kizu} 
which gives the complete one loop contribution including the
 gluino, the chargino and the neutralino exchanges against 
 which the signs of the relative contributions 
  of the various terms can be checked. 
 Because of the sensitive
		issue of the cancellation between the gluino and the 
		chargino exchange diagrams crucial to the analysis of this
		paper, we have redone the
		full one loop analysis of the EDM with the 
		gluino, the chargino and the neutralino exchanges.  We compare
		our results to those of Ref.~\cite{kizu,garis} in Appendix A. 
		In our analysis we have
		made the standard assumption of ignoring 
	      	all the generational mixing of quarks and of squarks.
	      	
	      		The outline of the paper is as follows: In Sec. 2
	      		we give the general features of the minimal
	      		supergravity unified model and discuss the new
	      	     CP violating phases it supports. In Sec.3 we give
	      	     our evaluation of the gluino, the chargino and the  
	      	     neutralino contributions to the electric dipole 
	      	     operator. In Sec.4 we display our evaluation of
	the chromoelectric  and of the purely gluonic operator 
	contributions.
	      	     Numerical analysis of results
	      	     and the phenomenon  of cancellation among the
	      	     various components is discussed in Sec.5. 
	      	     Conclusions are given in Sec.6. 
	      	     Diagonalization of the squark and of the chargino mass 
	      	     matrices
	      	     paying attention to the phases is given in Appendices
	      	      A and B along with a comparison of our results
                     with those of the previous analyses.

\section{N=1 Supergravity and CP Violating Phases }

	The  analysis of this paper is based on $N=1$
 supergravity grand unified theory in which
 supersymmetry is broken spontaneously via gravitational interactions in
  the hidden sector~\cite{cham,applied}, and the electro-weak symmetry is 
  broken via radiative effects. We assume 
that the grand unified theory (GUT) group $G$ breaks at the scale $M_G$
to the Standard Model gauge group, and after breaking of supersymmetry 
the tree effective theory can be characterized by the 
following symmetry breaking sector at the GUT scale 
\beqn
V_{SB}(0)=m_{0}^2 z_az^a + (m_{0} A_0W^{(3)}+B_0W^{(2)}+H.c.)+
 \frac{1}{2}m_{\frac{1}{2}}\bar\lambda_i\lambda_i
\eeqn
Here $W^{(2)}=\mu_0H_1H_2$, with $H_1$, $H_2$ being the two Higgs doublets,
$W^{(3)}$ is the superpotential cubic in the fields, 
 $m_0$ is the universal scalar mass, $A_0$ is the universal trilinear
coupling, $B_0$ is the universal bilinear coupling, and  $m_{\frac{1}{2}}$
is the universal gaugino mass.
 In general  $A_0$,  $B_0\mu_0$, $\mu_0$ and  
$ m_{\frac{1}{2}}$ are complex. However, not all the phases are 
physical. It is possible to 
remove the phases of $m_0, m_{\frac{1}{2}}$ and make $B_0\mu_0$    
real by redefinition of the fields and by
doing R-transformations on them.  We are then left with only two
independent phases at the GUT scale. One may choose one of these to be
 the phase of
 $\mu_0$ ($\theta_{\mu0}$), and the other to be the phase of  $A_0$
 ($\alpha_{A0}$).     
 Using renormalization group evolution one can evolve 
$V_{SB}(0)$ to low energy and one finds 
\beqn
V_{SB} & &={m_1^2|H_1|^2+m_2^2|H_2|^2 -
  [B \mu\epsilon_{ij} H_1^i H_2^j+H.c.]} \nonumber\\
& &
\hspace{4cm} +{M_{\tilde{Q}}^2[\tilde{u}_{L}^*\tilde{u}_{L}+
\tilde{d}_{L}^*\tilde{d}_{L}]+M_{\tilde{U}}^2 \tilde{u}_{R}^*\tilde{u}_{R}+
M_{\tilde{D}}^2 \tilde{d}_{R}^*\tilde{d}_{R}}\nonumber\\
& &
{+M_{\tilde{L}}^2[\tilde{\nu}_{e}^*\tilde{\nu}_{e}+
\tilde{e}_{L}^*\tilde{e}_{L}]
+M_{\tilde{E}}^2 \tilde{e}_{R}^*\tilde{e}_{R}}\nonumber\\
& &
{+\frac{g m_0}{\r2 m_W} \epsilon_{ij}[\frac{m_e A_e}{\cb} H_1^i \tilde{l}_{L}^j
\tilde{e}_{R}^* +\frac{m_d A_d}{\cb} H_1^i \tilde{q}_{L}^j
\tilde{d}_{R}^* -\frac{m_u A_u}{\sb} H_2^i \tilde{q}_{L}^j
\tilde{u}_{R}^*+H.c. ]}\nonumber\\
& &
{+\frac{1}{2}[\tilde{m}_3  \bar{\tilde{g}} \tilde{g} 
   +\m2 \bar{\tilde{W}}^a \tilde{W}^a+\m1 \bar{\tilde{B}} \tilde{B}]
   +\Delta V_{SB}}
\eeqn
where  ($\tilde{l}_{L}$, $\tilde{q}_{L}$) are the SU(2) (slepton, squark) 
doublets (the generation indices are suppressed), and $\Delta V_{SB}$ is
the one loop contribution to the effective potential\cite{coleman}.

 In our analysis  the electroweak
symmetry is broken by radiative effects which allows one to 
determine the magnitude of $\mu_0$ by fixing $M_Z$ and to find the 
magnitude of $B_0$ in terms of $\tb$=${\langle H_2 \rangle}\over
			{\langle H_1 \rangle}$.
In the  analysis we use one-loop renormalization group equations
 (RGEs) for the evolution of the soft SUSY breaking parameters and for
 the parameter $\mu$,  and two-loop RGEs for the gauge and Yukawa 
 couplings. The equations for the gauge and the Yukawa couplings, and 
 the diagonal 
elements of the sfermion masses and gaugino masses are such that they are 
entirely real, while the phase of $\mu$ doesn't run because 
it cancels out of the  one loop  renormalization group equation of $\mu$.
However, both the magnitudes and the phases of $A_i$  do evolve
and one has
\beq
\frac{d A_t}{d t}=-(\frac{16}{3}\tilde{\alpha}_3\frac{\tilde{m}_3}{m_0}
+3\tilde{\alpha}_2\frac{\tilde{m}_2}{m_0}+\frac{13}{15}\tilde{\alpha}_1
\frac{\tilde{m}_1}{m_0}+6 Y^t A_t + Y^b A_b ),
\eeq

\beq
\frac{d A_b}{d t}=-(\frac{16}{3} \tilde{\alpha}_3
\frac{\tilde{m}_3}{m_0}
+3\tilde{\alpha}_2\frac{\tilde{m}_2}{m_0}+\frac{7}{15}\tilde{\alpha}_1
\frac{\tilde{m}_1}{m_0}+ Y^t A_t +6 Y^b A_b + Y^{\tau} A_{\tau}),
\eeq

\beq
\frac{d A_{\tau}}{d t}=-(3 \tilde{\alpha}_2
\frac{\tilde{m}_2}{m_0}+
\frac{9}{5}\tilde{\alpha}_1
\frac{\tilde{m}_1}{m_0}+4 Y^{\tau} A_{\tau} + 3 Y^b A_b ),
\eeq
\beq
\frac{d A_u}{d t}=-(\frac{16}{3}\tilde{\alpha}_3\frac{\tilde{m}_3}{m_0}
+3\tilde{\alpha}_2\frac{\tilde{m}_2}{m_0}+\frac{13}{15}\tilde{\alpha}_1
\frac{\tilde{m}_1}{m_0}+3 Y^t A_t ),
\eeq

\beq
\frac{d A_d}{d t}=-(\frac{16}{3} \tilde{\alpha}_3
\frac{\tilde{m}_3}{m_0}
+3\tilde{\alpha}_2\frac{\tilde{m}_2}{m_0}+\frac{7}{15}\tilde{\alpha}_1
\frac{\tilde{m}_1}{m_0}+ 3Y^b A_b ),
\eeq
and
\beq
\frac{d A_e}{d t}=-(3 \tilde{\alpha}_2
\frac{\tilde{m}_2}{m_0}+
\frac{9}{5}\tilde{\alpha}_1
\frac{\tilde{m}_1}{m_0}+ 3 Y^b A_b ),
\eeq
where 
$\tilde{\alpha}_i$=${g_i^2}\over {(4\pi)^2}$,
$Y_{(u,d,e)}$=$h^{(u,d,e)^2}\over {(4\pi)^2}$, and where 
$g_i$ are the gauge couplings and $h^{(u,d,e)}$ are the Yukawa 
couplings, and $t=ln\frac{M_{G}^2}{Q^2}$.
The supergravity model with CP violation is then completely parametrized by 
just six quantities:
 $m_0,  m_{\frac{1}{2}},  A_0,  \tb,  \theta_{\mu0}$ and $\alpha_{A0}$.
There are 32 new particles in this model. Their masses   
 and interactions are  determined by the six parameters 
above. Thus the model is very predictive. 


\section{EDM Calculation}

	One defines the EDM of a spin-$\frac{1}{2}$ particle by 
 the effective lagrangian
\beq
{\cal L}_I=-\frac{i}{2} d_f \bar{\psi} \sigma_{\mu\nu} \gamma_5 \psi F^{\mu\nu}
\eeq
which in the non-relativistic limit gives
$ {\cal L}_I=d_f  \psi^{\dag}_A \vec{\sigma} . \vec{E} \psi_A$ where
$\psi_A$ is the large component of Dirac field.   
In renormalizable theories the effective lagrangian (9)
is induced at the  loop level if the theory contains a source of CP violation 
 at the tree level.      For a theory of fermion $\psi_f$
  interacting with other heavy
fermions $\psi_i$'s and heavy scalars $\phi_k$'s
with masses $m_i$, $m_k$ and charges $Q_i $, $Q_k$
 the interaction that contains CP violation in
general is given by
\beq
-{\cal L}_{int}=\sum_{ik} \bar{\psi}_f
                (K_{ik} \PL +L_{ik} \PR) \psi_i\phi_k+H.c.
\eeq    
Here  ${\cal L}$ violates CP invariance iff 
${\rm Im}(K_{ik} L_{ik}^*)\neq 0$.  
The one loop EDM of the fermion f in this  case is given by                     

\beq
\sum_{ik} \frac{m_i }{(4\pi)^2 m_k^2}{\rm Im}(K_{ik} L_{ik}^*)
        (Q_i A(\frac{m_i^2}{m_k^2})+Q_k    B(\frac{m_i^2}{m_k^2}))
\eeq
where A(r) and B(r) are defined by
\beq 
 A(r)=\frac{1}{2(1-r)^2}(3-r+\frac{2lnr}{1-r})
\eeq
and 
\beq
B(r)=\frac{1}{2(r-1)^2}(1+r+\frac{2rlnr}{1-r}),
\eeq
where  one  has
 charge conserved at the vertices, i.e., $Q_k=Q_f-Q_i$.
The loop diagrams corresponding to the term  $A$ is Fig.(1a) and to the
term B is Fig.(1b).\\

\noindent
\subsection{Gluino Contribution}

   The 	quark-squark-gluino interaction
    is given by ~\cite{applied}:
\beq
-{\cal L}_{q-\tilde{q}-\tilde{g}}=\r2 g_s T_{jk}^a \sum_{i=u,d} 
        (-\bar{q}_{i}^j \PL \tilde{g}_a \tilde{q}_{iR}^k +
	\bar{q}_{i}^j \PR \tilde{g}_a \tilde{q}_{iL}^k) + H.c. ,
\eeq
where $a=1-8$ are the gluino color indices, and $j,k=1-3$ are the quark and
squark color indices.
The scalar fields $\tilde{q}_L$ and $\tilde{q}_R$    
are in general linear combinations of the mass eigenstates which are given by
diagonalizing the squark $(mass)^2$  matrices 
for $\tilde{u}$ and $\tilde{d}$  at the electroweak scale~\cite{applied}:
\beq
M_{\tilde{u}}^2=\left(\matrix{M_{\tilde{Q}}^2+m{_u}^2+M_{z}^2(\frac{1}{2}-Q_u
\sin^2\theta_W)\cos2\beta & m_u(A_{u}^{*}m_0-\mu \cot\beta) \cr
   	          	m_u(A_{u} m_0-\mu^{*} \cot\beta) & M_{\tilde{U}
}^2+m{_u}^2+M_{z}^2 Q_u \sin^2\theta_W \cos2\beta}
		\right)
\eeq
and
\beq
M_{\tilde{d}}^2=\left(\matrix{M_{\tilde{Q}}^2+m{_d}^2-M_{z}^2(\frac{1}{2}+Q_d
\sin^2\theta_W)\cos2\beta & m_d(A_{d}^{*}m_0-\mu \tan\beta) \cr
                        m_d(A_{d} m_0-\mu^{*} \tan\beta) & M_{\tilde{D}
}^2+m{_d}^2+M_{z}^2 Q_d \sin^2\theta_W \cos2\beta}
                \right).
\eeq
where $Q_u=\frac{2}{3}$ and $Q_d=-\frac{1}{3}$.
  We note that  the $A_u$ and $A_d$ are not independent but evolve from
  the same common $A_0$ at the GUT scale. Further, $\theta_{\mu}$(the phase
  of $\mu$), $\alpha_{A_u}$(the phase of $A_u$),   and   $\alpha_{A_d}$
  (the phase of $A_d$) are related to just the two phases  
   $\theta_{\mu_0}$ and $\alpha_{A_0}$ at the GUT scale by renormalization 
   group evolution.
We diagonalize the squark matrices so that 
\beq
\tilde{q}_L=D_{q11} \tilde{q}_1 +D_{q12} \tilde{q}_2
\eeq
\beq
\tilde{q}_R=D_{q21} \tilde{q}_1 +D_{q22} \tilde{q}_2.
\eeq
where $\tilde{q}_1$ and  $\tilde{q}_2$ are the mass eigenstates. A more
detailed discussion of the diagonalization is given in Appendix A.

In terms of the mass eigenstates $\tilde{q}_1$ and  $\tilde{q}_2$ 
  the gluino contribution is given by 
\beq
{d_{q-gluino}^E}/{e}=-\frac{2\alpha_s}{3\pi} \sum_{k=1}^{2}
     {\rm Im}(\Gamma_{q}^{1k}) \frac{m_{\tilde{g}}}{M_{\tilde{q}_k}^2}
     Q_{\tilde{q}} {\rm B}(\frac{m_{\tilde{g}}^2}{M_{\tilde{q}_k}^2}),
\eeq
where $\Gamma_{q}^{1k}=D_{q2k} D_{q1k}^*$, 
  $\alpha_s$=${g_{s}^2}\over {4\pi}$, $m_{\tilde{g}}$ is the gluino mass,
  and $e$ is the positron charge.

\subsection{Neutralino Contribution}
In order to discuss the neutralino exchange contributions we first exhibit
the neutralino mass matrix:

\beq
M_{\chi^0}=\left(\matrix{\m1 & 0 & -\mz\sw\cb & \mz\sw\sb \cr
                 0  & \m2 & \mz\cw\cb & -\mz\cw\sb \cr
		 -\mz\sw\cb & \mz\cw\cb & 0 & -\mu \cr
		 \mz\sw\sb  & -\mz\cw\sb & -\mu & 0}
			\right).
\eeq
The matrix $M_{\chi^0}$ is a complex non hermitian and symmetric matrix, 
which can be  diagonalized  using a unitary matrix $X$ such that
\beq
X^T M_{\chi^0} X={\rm diag}(\mx1, \mx2, \mx3, \mx4).
\eeq
By rearranging the fermion-sfermion-neutralino interaction\cite{us},
the neutralino exchange contribution to the fermion EDM  is given by
\beq
{d_{f-neutralino}^E}/{e}=\frac{\AEM}{4\pi\sinW2}\sum_{k=1}^{2}\sum_{i=1}^{4}
{\rm Im}(\eta_{fik})
               \frac{\mxi}{M_{\tilde{f}k}^2} Q_{\tilde{f}}
		{\rm B}(\frac{\mxi^2}{M_{\tilde{f}k}^2}),
\eeq
where
\beqn
\eta_{fik} & &={[-\r2 \{\tan\theta_W (Q_f-T_{3f}) X_{1i}
  +T_{3f} X_{2i}\}D_{f1k}^*+
     \kappa_{f} X_{bi} D_{f2k}^*]}\nonumber\\
& &
\hspace{4cm} {(\r2 \tan\theta_W Q_f X_{1i} D_{f2k}
     -\kappa_{f} X_{bi} D_{f1k})}.
\eeqn
Here we have 
\beqn
\kappa_u=\frac{m_u}{\r2 m_W \sb}, 
 ~~\kappa_{d,e}=\frac{m_{d,e}}{\r2 m_W \cb}
\eeqn
where $b=3(4)$ for $T_{3f}=-\frac{1}{2}$($\frac{1}{2})$.

\subsection{Chargino Contribution}

To discuss the contribution of the chargino exchanges  we exhibit 
first the  chargino mass matrix  
\beq
M_C=\left(\matrix{\m2 & \r2 m_W \sb \cr
	\r2 m_W \cb & \mu }
            \right)
\eeq
This matrix can be diagonalized  by the 
biunitary  transformation 
\beq
U^* M_C V^{-1}={\rm diag}(\mc1, \mc2)
\eeq
where U and V are unitary matrices(see Appendix B).
 By looking at the fermion-sfermion-chargino
interaction  we find 
the chargino contribution to the EDMs for the up quark, the down quark
 and for the electron as follows 
\beq
{d_{u-chargino}^{E}}/{e}=\frac{-\AEM}{4\pi\sinW2}\sum_{k=1}^{2}\sum_{i=1}^{2}
      {\rm Im}(\Gamma_{uik})
               \frac{\mci}{M_{\tilde{d}k}^2} [Q_{\tilde{d}}
                {\rm B}(\frac{\mci^2}{M_{\tilde{d}k}^2})+
	(Q_u-Q_{\tilde{d}}) {\rm A}(\frac{\mci^2}{M_{\tilde{d}k}^2})],
\eeq

\beq
{d_{d-chargino}^{E}}/{e}=\frac{-\AEM}{4\pi\sinW2}\sum_{k=1}^{2}\sum_{i=1}^{2}
   {\rm Im}(\Gamma_{dik})
               \frac{\mci}{M_{\tilde{u}k}^2} [Q_{\tilde{u}}
                {\rm B}(\frac{\mci^2}{M_{\tilde{u}k}^2})+
        (Q_d-Q_{\tilde{u}}) {\rm A}(\frac{\mci^2}{M_{\tilde{u}k}^2})],
\eeq
and
\beq
{d_{e-chargino}^{E}}/{e}=\frac{\AEM}{4\pi\sinW2} \frac{\kappa_e}
    {m_{\tilde{\nu}e}^2} \sum_{i=1}^{2} \mci {\rm Im}(U_{i2}^* V_{i1}^*)
	{\rm A}(\frac{\mci^2}{m_{\tilde{\nu}e}^2})
\eeq
where
\beq
\Gamma_{uik}=\kappa_u V_{i2}^* D_{d1k} (U_{i1}^* D_{d1k}^*-
		\kappa_d U_{i2}^* D_{d2k}^*)
\eeq

\beq
\Gamma_{dik}=\kappa_d U_{i2}^* D_{u1k} (V_{i1}^* D_{u1k}^*-
                \kappa_u V_{i2}^* D_{u2k}^*).
\eeq
The sum total of  the gluino, the neutralino and the chargino contributions 
to the EDM gives us the total EDM. To obtain the 
  neutron EDM contribution from the electric dipole moment operator
we use the non-relativistic $SU(6)$ quark model which
  gives
\beq
d_n=\frac{1}{3}[4 d_d - d_u].
\eeq

The analysis of $d_n$ above is at  the electro-weak
scale and it must be evolved down to the hadronic scale 
via renormalization group
evolution  to give 
\beq
d_n^E=\eta^{E} d_n
\eeq
where $\eta^{E}$ is the QCD correction factor and we estimate it to be 1.53
in agreement with the analysis of ref.~\cite{arno2}

\section{The Chromoelectric and the CP Violating Purely Gluonic
Dimension Six  Operators}
The quark chromoelectric dipole moment is defined to be the factor $\tilde
d^C$
in the effective operator
\begin{equation}
{\cal L}_I=-\frac{i}{2}\tilde d^C \bar{q} \sigma_{\mu\nu} \gamma_5 T^{a} q
 G^{\mu\nu a}
 \end{equation}
 where $T^a$ are the generators of $SU(3)$. The gluonic dipole
moment $d^G$ is defined to be the factor in the
effective operator
\begin{equation}
{\cal L}_I=-\frac{1}{6}d^G f_{\alpha\beta\gamma}
G_{\alpha\mu\rho}G_{\beta\nu}^{\rho}G_{\gamma\lambda\sigma}
\epsilon^{\mu\nu\lambda\sigma}
\end{equation}
where $G_{\alpha\mu\nu}$ is the
 gluon field strength tensor, $f_{\alpha\beta\gamma}$
 are the Gell-Mann coefficients, and $\epsilon^{\mu\nu\lambda\sigma}$
is the totally antisymmetric tensor with $\epsilon^{0123}=+1$.
An analysis of these operators 
in minimal supergravity with two CP violating phases was given in
ref.\cite{us}.
We quote the results from that work here.  For the chromoelectric
dipole moment one has three contributions; from the  gluino exchange, 
from the neutralino exchange,  and from the chargino exchange. These are
given by\cite{us}
\beq
\tilde d_{q-gluino}^C=\frac{g_s\alpha_s}{4\pi} \sum_{k=1}^{2}
     {\rm Im}(\Gamma_{q}^{1k}) \frac{m_{\tilde{g}}}{M_{\tilde{q}_k}^2}
      {\rm C}(\frac{m_{\tilde{g}}^2}{M_{\tilde{q}_k}^2}),
\eeq
where 
\beq
C(r)=\frac{1}{6(r-1)^2}(10r-26+\frac{2rlnr}{1-r}-\frac{18lnr}{1-r}),
\eeq

\beq
\tilde d_{q-neutralino}^C=\frac{g_s g^2}{16\pi^2}\sum_{k=1}^{2}\sum_{i=1}^{4}
{\rm Im}(\eta_{qik})
               \frac{\mxi}{M_{\tilde{q}k}^2}
                {\rm B}(\frac{\mxi^2}{M_{\tilde{q}k}^2}),
\eeq
and 
\beq
\tilde d_{q-chargino}^C=\frac{-g^2 g_s}{16\pi^2}\sum_{k=1}^{2}\sum_{i=1}^{2}
      {\rm Im}(\Gamma_{qik})
               \frac{\mci}{M_{\tilde{q}k}^2}
                {\rm B}(\frac{\mci^2}{M_{\tilde{q}k}^2}).
\eeq

The contribution to the EDM of the quarks can be computed using the naive
dimensional analysis~\cite{georgi} which gives
\beq
d^C_q=\frac{e}{4\pi} \tilde d^C_{q} \eta^C
\eeq
where $\eta^C$ is the QCD correction factor for the color dipole operator.

For the CP violating dimension six operator 
\beq
d^G=-3\alpha_s m_t (\frac{g_s}{4\pi})^3
{\rm Im} (\Gamma_{t}^{12})\frac{z_1-z_2}{m_{\tilde{g}}^3}
{\rm H}(z_1,z_2,z_t)
\eeq
where
\beq
z_{\alpha}=(\frac{M_{\tilde{t}\alpha}}{m_{\tilde{g}}})^2,
z_t=(\frac{m_t}{m_{\tilde{g}}})^2
\eeq
The contribution to $d_n$ from $d^G$ can be estimated by the naive dimensional
analysis\cite{georgi} which gives
\beq
d_{n}^G=\frac{eM}{4\pi} d^G \eta^G
\eeq
where $M$ is the chiral symmetry breaking scale and has  the numerical value
1.19 GeV, and $\eta^G$ is the renormalization group evolution factor
of the dimension
six operator from the electroweak scale down to the hadronic scale.
We estimate that $\eta^C$ $\approx$ $\eta^G$ $\sim 3.4$ in agreement 
with the analysis of ref.~\cite{arno2}. To get the contributions of the 
chromoelectric and dimension six operators to EDM we used the reduced
coupling constant and the naive dimensional analysis. There is another
way of estimating this contribution for the chromoelectric operator
and that is using QCD sum rules~\cite{khat}. The use of QCD sum rules
rather than of the naive dimensional method would not change the conclusions
of this paper.

\section{EDM Analysis}

As already stated while the EDM of the neutron has been analysed in
many works, most of the previous analyses  
 have been carried out within MSSM. 
 Our analysis here is in the framework of N=1 supergravity and
 we use radiative breaking 
of the electro-weak symmetry including one loop effective potential 
terms\cite{coleman} to analyse the 
 EDMs in the six dimensional parameter space of the theory given by
 $m_0,  m_{\frac{1}{2}},  A_0,  \tb,  \theta_{\mu0}$ and
$\alpha_{A0}$. 
The constraints imposed on the radiative electro-weak symmetry 
breaking include
imposition of color and charge conservation, experimental lower limit 
constraints on the sparticle masses from LEP, CDF and DO, and the
experimental constraints on $b\rightarrow s+\gamma$ from CLEO\cite{ammar}.
 (Details of the analysis are similar to those of Ref.\cite{swieca}.).
As mentioned in Sec.1  in most of the previous analyses in the
literature the effects of the chromo-electric and of the purely gluonic 
operators have been assumed small and ignored. As shown in ref.\cite{us}
this is an erroneous assumption as the relative contributions of the 
electric, of the chromo-electric , and of the purely gluonic operators are 
highly model dependent and their ratios can sharply change as one moves
in the six dimensional parameter space of the model.  In fact, it was shown
in ref.\cite{us} that contrary to the assumptions generally made the 
contributions of the chromoelectric and of the purely gluonic operators 
can be comparable to and may even exceed the contribution of the electric dipole 
term. Because of the significant contribution  that the chromo-electric and
the purely gluonic operators can make to the neutron EDM we include  in our 
analysis all the three contributions, i.e., the electric, the
chromo-electric and the purely gluonic operator contributions. 
 However, we do not include in the analysis the effects induced by  the 
 phase in the Kobayashi-Maskawa(KM) mass matrix in the renormalization 
group evolution of the SUSY phases, since these induced effects are 
known to be very small\cite{inui,posp,berto}.

One of the important phenomenon we find in our analysis is the 
possibility of  destructive interference between the gluino and the
chargino exchange diagrams for the electric dipole and for the 
chromoelectric  terms. This generally happens when the signs of phases
of $\theta_{\mu_0}$ and $\alpha_{A_0}$ are opposite. In this case there
is also a destructive interference between the $\mu$ and the $A_t$
terms in the purely gluonic part. In addition to the above one also
 finds a further cancellation among the electric, the chromoelectric and
 the purely gluonic parts.  
 Constraints on the theoretical analyses are provided
by the experimental upper limits on the EDMs. 
 For the neutron the current experimental
limit is \cite{altra}
\beq
d_n < 1.1\times 10^{-25} ecm
\eeq
and for the electron the limit is \cite{commins}
\beq
d_e < 4.3\times 10^{-27} ecm
\eeq
For the muon the  current experimental upper limit is 
$d_{\mu}<1.1\times 10^{-18}$ ecm\cite{bailey}(at 95$\%$ CL). 
This limit may improve by 
up to four orders of magnitude in a new proposed experiment at the
Brookhaven National Laboratory\cite{bnl}. 
However, the constraints on the supergravity parameter space 
from the current limits on the neutron EDM and on the 
electron EDM
are already much stronger than what might emerge from the improved
muon EDM experiment. For this reason we  focus in our analysis 
on the constraints coming from the neutron EDM and from the electron
EDM. However, we shall sometimes also display  the muon EDM  for comparison 
along  with the neutron and the electron EDM.

We begin our discussion with a comparison between
the electron and neutron EDMs constraints on the
two basic parameters of the theory, i.e., $m_0$ and
$m_{1/2}$. As may be seen from Fig.(2a) the electron EDM falls off with 
increasing $m_0$. This behavior is easily understood  from  Eqs.(22) and
(29) since as $m_0$ increases $A(r)/M_{\tilde{f}}^2$ and 
$B(r)/M_{\tilde{f}}^2$ decrease. Using the experimental upper limit of Eq.(45) 
in Fig.(2a) one finds  for  $|A_0|=$ 1.0, $\tb=$ 3.0 and
$\alpha_{A0}= \theta_{\mu0}=\frac{\pi}{10}$  the
 following constraints on $m_0$: $m_0>1320$  GeV for $m_{1/2}=700$ GeV, 
 $m_0>1420$ GeV for $m_{1/2}=600$ GeV, and $m_0>1520$ GeV for $m_{1/2}=500$ 
 GeV. A similar analysis  holds for Fig.(2b). Here using the experimental upper 
 limit on the neutron EDM of Eq.(44)  and for the same above parameters 
one finds: $m_0>2500$ GeV for $m_{1/2}
 =700$ GeV, $m_0>2680$ GeV for $m_{1/2}=600$ GeV and $m_0>2840$ GeV for 
 $m_{1/2}=500$ GeV. Thus in this region of the parameter space 
 the upper limit on  the neutron 
 EDM rather than the upper limit on the electron EDM is the more
 severe constraint on $m_0$. The dependence of 
 $d_e$ and $d_n$ on $m_{1/2}$ is 
 displayed in Figs.(2c) and (2d). 
 The broad maxima for small  $m_{\frac{1}{2}}$  in these graphs
arise from an interplay between the  factors $\mci$ ,$\mxi$ and 
 $m_{\tilde g}$ which increase as  $m_{\frac{1}{2}}$ increases, and 
 the functions $A(r)$, $B(r)$ and $C(r)$ which decrease as  $m_{\frac{1}{2}}$
 increases. 
 By carrying out the same analysis as for the $m_0$ dependence, 
 one finds  here also that  the experimental upper limit constraint for
 the neutron EDM is a more severe constraint than the one for the electron 
 EDM. 

The dependence of the EDMs on $\theta_{\mu_0}$ in displayed in Fig.(2e) 
and on tan$\beta$ in Fig.(2f). The conventional fine tuning problem 
can be understood from the analysis of Fig.(2e) where the phase 
$\theta_{\mu0}$ must lie in a very small corridor around the origin to
satisfy the current experimental constraints on the neutron EDM. 
Fig.(2f) shows that the  EDMs are an increasing function of tan$\beta$.
This behavior can be understood easily for the electron and for the muon
EDM since these involve a factor of $1/cos\beta$ which increases
 as tan$\beta$ increases. For  the neutron EDM case, there are contributions 
 from
both the up quark and from the down quark with different tan$\beta$ dependences.
However, the down quark contribution dominates 
and as Fig.(2f) shows the  neutron EDM is still an increasing
function of tan$\beta$. 

In the analysis thus far we did not take advantage of the
two independent phases.
To give a comparison of the results arising in the two cases we consider
first the case of Fig.(3a) where the 
signs of $\alpha_{A_0}$ and $\theta_{\mu_0}$ are both positive. 
Here we find that there are no large internal cancellations within  
 the various components , $d_n^E$, $d_n^C$, and
$d_n^G$  so these functions do not show any rapidly varying behavior.
However, $d_n^E$ and $d_n^C$ in this case are negative while 
$d_n^G$ is positive over the entire $|A_0|$ region and there is  a 
cancellation among them. In the region
of $|A_0|\leq 2.5$ the cancellation is rather small because $d_n^G$ is
relatively small. However, the cancellation becomes more 
significant for $|A_0|>2.5$
leading to a dip in the total $d_n$ in this region as seen in Fig.(3a).

We consider next the case in Fig.(3b) when the sign of $\alpha_{A_0}$ is 
switched. Here each of the individual components $d_n^E$, 
$d_n^C$, and $d_n^G$ shows a destructive interference giving rise 
to  sharp minima  as a  function of  $|A_0|$. These 
minima can be understood as follows: For the case of $d_n^E$ and
$d_n^C$ the minima arise as a consequence of destructive interference
between the gluino exchange and the chargino exchange in the one loop diagrams.
This illustrates what we have said previously
 that the  chargino exchange contributions are as important as the 
gluino exchange contributions  and  should be included in the analysis 
contrary to
what is often done in the literature.  The  minimum
in  $d_n^G$  in Fig.(3b) has a different origin. It can be understood by 
examining the expression 
 \beq
{\rm Im}(\Gamma_{t}^{12})=\frac{-m_t}{(M_{\tilde{t}1}^2-M_{\tilde{t}2}^2)}
        (m_0 |A_t| \sin \alpha_{t} + |\mu| \sin \theta_{\mu} \cot\beta),
\eeq
where $\theta_{\mu}$ and  $\alpha_{t}$ are the values of
$\theta_{\mu_0}$ and of $\alpha_{A_t}$ at the electro-weak scale. From
Eq.(46) we see that the magnitude of ${\rm Im}(\Gamma_{t}^{12})$
 depends on the relative sign and the relative 
magnitudes of $\theta_{\mu}$ and of $\alpha_{t}$. Thus a cancellation occurs 
between the $A_t$ and the $\mu$ terms  when
 $\theta_{\mu}$ and  $\alpha_{t}$ have opposite signs leading to  a 
 sharp minimum in $d_n^G$ as a function of $|A_0|$.
Now each of the three terms $d_n^E$, $d_n^C$, and $d_n^G$, switch 
sign as they pass their zero values. Thus $d_n^E$ and  $d_n^C$ 
are negative below their respective minima and become positive after
crossing them, while  $d_n^G$ is positive below the minimum and 
becomes negative after crossing it. This complex structure now gives
rise to two distinct minima in the albegraic sum of the three terms, i.e., 
in the total $d_n$ as may be seen in Fig.(3b). 
We pause here to note that for the case when there is destructive 
interference between the gluino and the chargino case, and a 
 further cancellation among the electric, the chromoelectric and the
purely gluonic terms as is the case for Fig.(3b), one finds  a drastic 
reduction in the magnitude of $d_n$ often by a factor 
 $O(10-10^3)$.

 In Fig.(3c) we give  a plot of the 
EDMs of the electron, the muon and the neutron as a function of $m_{1/2}$ 
showing the cases when $\alpha_{A_0}$ is positive and 
when $\alpha_{A_0}$ is negative.
  Here for the case when $\alpha_{A_0}$ is positive 
the neutron EDM is large enough that
it  violates the current 
experimental bound in the entire range of $m_{1/2}\leq 750$ GeV.
However, for the case when $\alpha_{A_0}$ is negative 
the neutron EDM lies
 below the experimental upper  limit for  $m_{1/2}\geq 300$ GeV. 
The large disparity between the magnitudes of the neutron EDM for the    
$\alpha_{A_0}$ positive  case vs  for the 
$\alpha_{A_0}$  negative case can shift the balance
between which of  the two experimental constraints, i.e.,
the experimental upper limit constraint on the neutron EDM or the 
experimental upper limit constraint on the 
electron EDM, is  the more stringent one.
It can be seen that for  the case of  constructive interference the
experimental upper limit constraint on the neutron EDM is generally the more
stringent one while for 
the case of destructive interference involving large
cancellation it is the experimental constraint on the electron EDM 
which  may be the  more stringent constraint.
 We shall exhibit this effect further in the analysis of Fig.(3e).

The destructive interference between the different contributions
  exhibited in Figs. (3a) - (3c) is not
an isolated phenomenon but rather a common occurrence in a 
large part of the parameter space. Thus, cancellations
 occur naturally over the entire parameter space with the appropriate 
 choice for the relative sign  
of $\theta_{\mu}$ and $\alpha_{A0}$. Further, these cancellations 
  can become exceptionally large in certain regions of the parameter space.
 An example of this effect already occurs in the analysis of Figs.(3b) 
 and (3c). Similar cancellations also appear in other regions of the
 parameter space.  
In Fig.(3d) the effect of cancellations in $d_n$ is shown as a function of 
$m_0$ for three sets of input data for the case when
 $\theta_{\mu}$ and $\alpha_{A0}$ have opposite signs. In each case there 
 are large cancellations which lead to the appearance of minima.   
	Aside from the reduction of the EDMs by  cancellations, there
	are regions of the parameter space where kinematical
	suppressions occur.
An example of this is the reduction of the EDMs when 
tan$\beta$ becomes small as may be seen in Fig.(2f).
 A kinematical suppression of the EDMs can also occur  if 
 $m_{\frac{1}{2}}/m_0<<1$. As one can see from Eqs.(19) and (29) that in 
 this case  the quark and the lepton EDMs are  kinematically suppressed.
 A suppression of this type appears to arise in  
 supersymmetric models with anomalous U(1) mediated supersymmetry
 breaking\cite{dvali}.

   Finally, we exhibit in Fig.(3e) the excluded regions in the
          $m_0$-$m_{1/2}$ plane under the  
         constraints given by the current experimental upper limits on
         $d_e$ and $d_n$. The regions
	between the axes and the curves are the excluded regions in 
	Fig.(3e). 
	The analysis of Fig.(3e) shows the dramatic effect of the 
	destructive interference on the allowed and the disallowed region in
	 the mass plot. One finds that destructive interference softens 
	 significantly the stringent constraints on $m_0$  and $m_{1/2}$.
	 Thus the excluded region in the $m_0-m_{1/2}$ plane 
	  for the destructive interference case is much smaller than
		  for the constructive interference case. 
	The analysis of Fig.(3e) illustrates another interesting
	phenomenon alluded to earlier. One finds from Fig.(3e) that  	 
	for the  constructive interference case	the $d_n$ 
	experimental constraint is the more severe one as it eliminates a 
	larger part of the parameter space, while for the destructive interference
	case the $d_e$ experimental constraint is the more severe one 
	as it excludes a larger part of the  parameter space in the  
	$m_0$-$m_{1/2}$  plane in this case.  
	
	In the above we have discussed cancellations which can result in a 
	drastic reduction for the case of the neutron edm. There  can also
	be cancellations for the case of the electron edm between the 
	chargino and the neutralino contributions. For comparable sizes of
	$\theta_{\mu}$ and $\alpha_{A_0}$, the chargino contribution is 
	much larger than the neutralino contribution and cancellation is not
	very effective. However, more significant cancellations can occur
	for very small values of $\theta_{\mu}$ and for moderate values 
	of $\alpha_{A_0}$ since in this case the contribution from the
	chargino exchange and the neutralino exchange become comparable.

\section{Conclusion}

	In this paper we have presented an
analysis of the EDM of the neutron and of the charged leptons 
within the framework of 
supergravity
grand unification under the constraint of radiative breaking of the 
electroweak
symmetry. All the supersymmetric one-loop contributions to the EDMs 
 were analyzed taking care of their relative signs.
 For the neutron we considered also the 
contributions from the chromoelectric and from the purely gluonic operators.
One finds that
 there exist significant regions of the parameter space where cancellations
  occur among the
different contributions for the case of the neutron electric dipole
 moment. In these regions the neutron EDM undergoes a 
 significant reduction and the current experimental limits are 
 consistent in these regions with CP violating phases which are not
 too small and with a SUSY mass spectrum which satisfies the naturalness
 constraint. One also finds that regions of the parameter space exist 
 where the destructive interference between the different components
 can reduce the magnitude of the neutron EDM even below the 
 magnitude of the electron EDM.

    The nature of interference, i.e.,
 constructive vs destructive, for the neutron EDM determines which of the two 
 experimental upper limit constraints, i.e., the upper limit on the neutron
 EDM, or the upper limit on the electron EDM, will constitute the 
 more stringent constraint. For the case of constructive interference for
 $d_n$, it is the experimental upper limit on $d_n$ itself  
  which is found to be generally more stringent constraint than the upper
 limit constraint on $d_e$. However, for the
 destructive interference case for $d_n$, one finds that it is 
 generally the upper limit constraint
on $d_e$ which becomes the more stringent constraint.

	As mentioned already the 
	previously known mechanisms for the suppression of the neutron
	EDM in SUSY
	theories consist of suppression either by
	a fine tuning using  small phases or by a choice of  a heavy SUSY 
	spectrum.
	We have pointed out a third possibility, i.e., that of 
	internal cancellations, which naturally suppress the neutron EDM
	without the necessity of either having very small phases or
	having an excessively heavy SUSY spectrum. 
	The cancellations that occur do not constitute a fine tuning. 
	Rather, one finds that such cancellations occur naturally over a 
	large part of the parameter space, and in some regions the
	cancellations become exceptionally large. 
	This result has important implications for the discovery
	of supersymmetric particles. 
	With the cancellation mechanism the  SUSY spectrum  within the current  
	naturalness limits can be consistent with the present EDM 
	experimental constraints without the finetuning of phases, and
	such a spectrum should still be within reach of 
	the LHC. 
	At the same time one also expects that if SUSY phases are indeed
	$O(1-10^{-1})$ and the SUSY spectrum lies in the usual naturalness limit
	 of O(1) TeV, then with the suppression of the neutron EMD  via the 
	 cancellation mechanism  
	 the neutron and the electron EDMs 
	should become visible with improvements 
	 of O(10) in the sensitivity of the EDM experiments.
	 Finally we point out that although our analysis has been done in
	 the framework of supergravity unification with soft SUSY breaking 
	 sector parametrized by six parameters (including two CP violating
	 phases), the mechanism of internal cancellations pointed out in this
	 paper which can suppresss the edms should be applicable to a 
	 wider class of models such as models with non-universal soft
	 SUSY breaking. \\
\section{Acknowledgements} 

 This research was supported in part by NSF grant 
PHY-96020274.

\section{Appendix A}

	The squark$(mass)^2$ matrix   
\beq
M_{\tilde{q}}^2=\left(\matrix{M_{\tilde{q}11}^2 & M_{\tilde{q}12}^2  \cr
			 M_{\tilde{q}21}^2 & M_{\tilde{q}22}^2}
				\right),
\eeq
is hermitian and can be diagonalized by the unitary transformation
\beq
D_{q}^\dagger M_{\tilde{q}}^2 D_q={\rm diag}(M_{\tilde{q}1}^2,
              M_{\tilde{q}2}^2)
\eeq
where one parametrizes $D_q$ so that 
\beq
D_q=\left(\matrix{\cos \frac{\theta_q}{2} 
           & -\sin \frac{\theta_q}{2} e^{-i\phi_{q}} \cr
	   \sin \frac{\theta_q}{2} e^{i\phi_{q}}
		&\cos \frac{\theta_q}{2}}
		\right),
\eeq
Here $ M_{\tilde{q}21}^2=|M_{\tilde{q}21}^2| e^{i\phi_{q}}$
and we choose the range of $\theta_q$ so that 
${{-\pi}\over {2}} \leq  \theta_q \leq {{\pi}\over {2}}$ where 
$\tan \theta_q=
\frac{2|M_{\tilde{q}21}^2|}{M_{\tilde{q}11}^2-M_{\tilde{q}22}^2}$.  
	The eigenvalues $M_{\tilde{q}1}^2$ and $M_{\tilde{q}2}^2$    
can be determined directly from Eq.(48) or from the roots 
\beq
M_{\tilde{q}(1)(2)}^2=\frac{1}{2} (M_{\tilde{q}11}^2+M_{\tilde{q}22}^2)
	(+)(-)\frac{1}{2}[(M_{\tilde{q}11}^2-M_{\tilde{q}22}^2)^2 +
		4|M_{\tilde{q}21}^2|^2]^{\frac{1}{2}}.
\eeq
The (+) in Eq.(50) corresponds to choosing the structure of  the matrix
$M_{\tilde{q}}^2$ so that for 
$M_{\tilde{q}11}^2>M_{\tilde{q}22}^2$
 one has $M_{\tilde{q}1}^2>M_{\tilde{q}2}^2$	
	and vice versa. For our choice of the $\theta_q$ range one has
\beq
\tan \theta_q=\frac{2 m_{q} |A_q m_0 - \mu^*
R_q|}{M_{\tilde{q}11}^2-M_{\tilde{q}22}^2}
\eeq
where $R_u=\cob$ and $R_d=\tb$. Further
\beq
\sin\phi_q=\frac{m_0 |A_q| \sin \alpha_q+ |\mu| \sin \theta_{\mu} R_q}
        {|m_0 A_q  - \mu^* R_q|}. 
\eeq
Using the above we get 
\beq
{\rm Im}(\Gamma_{q}^{11})=-{\rm Im}(\Gamma_{q}^{12})=
	\frac{1}{2} \sin \phi_{q} \sin \theta_q
\eeq
where 
\beq
\sin \theta_q=\pm \frac{2 m_{q} |A_q m_0 - \mu^*
R_q|}{|M_{\tilde{q}1}^2-M_{\tilde{q}2}^2|}
\eeq
the $[+(-)]$ in Eq.(54) depends on whether
$M_{\tilde{q}11}^2-M_{\tilde{q}22}^2$ is [$>0(<0)$]. 
Thus Eq.(53) gives

\beq
{\rm Im}(\Gamma_{q}^{11})=\frac{m_q}{M_{\tilde{q}1}^2-M_{\tilde{q}2}^2}
        (m_0 |A_q| \sin \alpha_q + |\mu| \sin \theta_{\mu} R_q),
\eeq
which holds quite generally, i.e., for the case where 
 $M_{\tilde{q}1}^2>M_{\tilde{q}2}^2$ and for the case where 
 $M_{\tilde{q}1}^2<M_{\tilde{q}2}^2$.
Thus the gluino contribution to the EDM of the quark is given by 
\beq
{d_{q-gluino}^E}/{e}=\frac{-2 \alpha_{s}}{3 \pi}  m_{\tilde{g}}Q_{\tilde{q}} 
{\rm Im}(\Gamma_{q}^{11}) [\frac{1}{M_{\tilde{q}1}^2}
 {\rm B}(\frac{m_{\tilde{g}}^2}{M_{\tilde{q}1}^2}) -\frac{1}{M_{\tilde{q}2}^2}
{\rm B}(\frac{m_{\tilde{g}}^2}{M_{\tilde{q}2}^2})].
\eeq
One may expand  the right hand side of Eq.(56) around 
the  average squark mass. Defining
$M_{\tilde q}^2$=($M_{\tilde q_1}^2$+$M_{\tilde q_2}^2$)/2,
and expanding in the difference 
($M_{\tilde q_1}^2$-$M_{\tilde q_2}^2$),one finds in the lowest 
approximation

\begin{eqnarray}
{d_{q-gluino}^E}/{e}\simeq 
\frac{2 \alpha_{s}}{3 \pi}  m_{\tilde{g}}Q_{\tilde{q}} 
\frac{m_q}{M_{\tilde q}^4} 
  (m_0 |A_q| \sin \alpha_q + |\mu| \sin \theta_{\mu} R_q)\nonumber\\
  (B(\frac{m_{\tilde g}^2}{M_{\tilde q}^2}) 
  + \frac{m_{\tilde g}^2}{M_{\tilde q}^2} 
  D(\frac{m_{\tilde g}^2}{M_{\tilde q}^2})) 
\end{eqnarray}
where D(r) is given by
\beq
D(r)=\frac{1}{2(1-r)^3}(5+r+2lnr+\frac{6rlnr}{(1-r)}).
\eeq

As mentioned  already currently there  is some confusion in the literature 
regarding the  sign of the gluino  contribution  to the electric dipole 
 operator\cite{kizu,garis}.
We first compare our results with those of ref.\cite{garis}. The analysis
of \cite{garis} corresponds to neglecting the D term in  Eq.(57) 
and using $d_n\simeq \frac{4}{3}d_d$ which gives 

\beq
\frac{d_{n}}{e}\simeq \frac{-8 \alpha_{s}}{27 \pi} m_{\tilde{g}} m_{d}
 [\frac{m_0 |A_d| \sin \alpha_d+ |\mu| \sin \theta_{\mu} \tb}
{M_{\tilde{d}}^4}]
{\rm B}(\frac{m_{\tilde{g}}^2}{M_{\tilde{d}}^2}).
\eeq
This result then agrees both in sign and in
magnitude with Eq.(3) of ref.\cite{garis}. To compare with 
the result of ref.\cite{kizu} we switch the sign of 
the $m_H$ term in their Eq.(6) (see e.g., ref.~\cite{haber})
 and find that our 
Eq.(19) differs from Eq.(14) of ref.\cite{kizu} by an overall minus sign.
A comparison of the chargino and the neutralino contributions with those
of ref.\cite{kizu} is more involved since the 
 chargino (and the neutralino) mass matrices are  different 
  in the two works. 
This difference arises because 
after  $ SU(2)_L\times U(1)_Y $ breaking to $U(1)_{EM}$,
 the authors of ref.\cite{kizu} expand the potential 
around the VEV so that  $H_i \rightarrow H_i-{\langle H_i \rangle}$
 instead of $H_i \rightarrow H_i+{\langle H_i \rangle}$
 and they use in the chargino case
$M_{C}^T$ instead of $M_C$  as is conventionally done\cite{haber}. 
Thus  to compare with their expressions we have to do the 
transformation:
$V_{ij}\rightarrow C_{Rji}$, $ U_{ij}^*\rightarrow C_{Lji}$,
$D \rightarrow S$ and
$X \rightarrow N$. After that, and assuming the conventional expansion
around the VEV, we go to their convention by  the transformation
$\kappa_f \rightarrow -\kappa_f$ to
find that we have the same overall
 sign in the case of the chargino exchange 
 but the sign of the $\kappa_{f'}$ term in
 the brackets in their Eq.(10) should be positive. 
 In the case
of neutralino exchange our result differs from their Eq.(12)  
by an overall sign
and further we find that the second term in the last bracket of 
their Eq.(13) (the term which begins with -$\kappa_f$) should have 
an opposite sign.

\section{Appendix B}

The chargino matrix $M_C$ is
 not hermitian, not symmetric and not real
because $\mu$
   is complex. $M_C$ is diagonalized using the biunitary transformation
\beq
U'^{*} M_C V^{-1}=M_D
\eeq
where $U'$ and $V$ are hermitian and $M_D$ is a diagonal matrix but not
yet real. $U'$ and V satisfy the relation
\beq
V (M_C^{\dagger} M_C) V^{-1}={\rm diag}(|\mc1|^2, |\mc2|^2)
=U'^{*} (M_C M_C^{\dagger}) (U'^*)^{-1}
\eeq
We may parametrize $U'$ so that 
\beq
U'=\left(\matrix{\cos \frac{\theta_1}{2}
           & \sin \frac{\theta_1}{2} e^{i\phi_{1}} \cr
           -\sin \frac{\theta_1}{2} e^{-i\phi_{1}}
                &\cos \frac{\theta_1}{2}}
                \right),
\eeq
where
\beq
\tan\theta_1=\frac{2\r2m_W[\m2^2\cos^{2}\beta+|\mu|^2\sin^2\beta
+|\mu|\m2\sin2\beta \cos\theta_{\mu}]^{\frac{1}{2}}}{\m2^2-|\mu|^2
-2m_{W}^2\cos2\beta}
\eeq
and
\beq
\tan\phi_1=\frac{|\mu|\sin\theta_{\mu}\sin\beta}{\m2 \cos\beta +|\mu|
\cos\theta_{\mu} \sin\beta}
\eeq
Similarly we parametrize V so that     
\beq
V=\left(\matrix{\cos \frac{\theta_2}{2}
           & \sin \frac{\theta_2}{2} e^{-i\phi_{2}} \cr
           -\sin \frac{\theta_2}{2} e^{i\phi_{2}}
                &\cos \frac{\theta_2}{2}}
                \right),
\eeq
where 
\beq
\tan\theta_2=\frac{2\r2m_W[\m2^2\sin^{2}\beta+|\mu|^2\cos^2\beta
+|\mu|\m2\sin2\beta \cos\theta_{\mu}]^{\frac{1}{2}}}{\m2^2-|\mu|^2
+2m_{W}^2\cos2\beta}
\eeq
and
\beq
\tan\phi_2=\frac{-|\mu|\sin\theta_{\mu}\cos\beta}{\m2 \sin\beta +|\mu|
\cos\theta_{\mu} \cos\beta}.
\eeq
We wish to choose the phases of $U'$ and $V$ so that the elements of $M_D$
will be positive. Thus we define  $U=H \times U'$ where
\beq
H=\left(\matrix{e^{i\gamma_1} & 0 \cr
            0 &e^{i\gamma_2}}
                \right),
\eeq   
such that
\beq
U^{*} M_{C} V^{-1}= \left(\matrix{|\mc1| & 0 \cr
            0 &|\mc2|}
                \right),
\eeq
where $\gamma_1$ and $\gamma_2$ are the phases of the diagonal elements in 
 Eq.(60). Our 
 choice of the signs  and the roots is such that 
\beqn
M_{(\mc1)(\mc2)}^2 & &={\frac{1}{2} [\m2^2+|\mu|^2+2m_{W}^2](+)(-)}\nonumber\\
& &
{\frac{1}{2}[(\m2^2-|\mu|^2)^2+4m_{W}^4 \cos^{2}2\beta+4m_{W}^2}\nonumber\\
& &
(\m2^2+|\mu|^2+2\m2 |\mu| \cos\theta_{\mu} \sin2\beta)]^\frac{1}{2}
\eeqn
where the  sign  chosen is such that  $\mc1<\mc2$ if
\beq
\m2^2<|\mu|^2+2m_{W}^2 \cos2\beta.
\eeq

	For the neutralino matrix, the eigenvalues and the diagonalizing
matrix $X$ must be estimated numerically.

\newpage	
\begin{center}
{\large\bf Figure Captions}
\end{center}
\vspace{0.5cm}
\noindent

\noindent
Fig.~(1a): One loop  diagram contributing to the electric dilpole operator where
the external photon line ends on an exchanged chargino line labelled by 
$\tilde \chi_i^+$ in the loop.\\

\noindent
Fig.~(1b): One loop diagram contributing to the electric dipole operator 
 where the external photon line ends on an exchanged squark (slepton) line
 represented by $\tilde q_k(\tilde l_k)$ on the internal line. \\

\noindent
Fig.~(2a): Plot  of the magnitude of the electron EDM as a function 
of $m_0$ when $|A_0|=$ 1.0, $\tb=$ 3.0 and
$\alpha_{A0}= \theta_{\mu0}=\frac{\pi}{10}$ for different values of 
$m_{1/2}$.  The dotted curve is  for
$m_{1/2}=500$ GeV, the solid curve for $m_{1/2}=600$ GeV,
and the dashed curve is for $m_{1/2}=700$ GeV.\\

\noindent 
Fig.~(2b): Plot of the magnitude of the neutron EDM as a function of $m_0$
for the same parameters as in Fig.~2(a).\\

\noindent
Fig.~(2c): Plot  of the magnitude of the electron EDM as a function 
of $m_{1/2}$ when $|A_0|=$ 1.0, $\tb=$ 3.0 and
$\alpha_{A0}= \theta_{\mu0}=\frac{\pi}{10}$ for different values of 
$m_0$.  The dotted curve is  for
$m_0=500$ GeV, the solid curve for $m_0=1000$ GeV
and the dashed curve is for $m_0=1500$ GeV.\\

\noindent
Fig.~(2d): Plot of the magnitude of the neutron EDM as a function of $m_{1/2}$
for the same parameters as in Fig.~2(c).\\

\noindent
Fig.~(2e): Plot of the magnitudes of the neutron, the electron and the muon
EDMs as a function of $\theta_{\mu_0}$ for the case when $|A_0|=$ 1.0, 
$\tb=$ 3.0, $\alpha_{A0}= \frac{\pi}{20}$, 
$m_0=1000$ GeV and $m_{1/2}=500$ GeV.\\

\noindent
Fig.~(2f): Plot of the magnitudes of the neutron, the electron and the muon
EDMs  as a function of tan$\beta$ for the case when 
$|A_0|=$ 1.0, $\alpha_{A0}= \theta_{\mu0}=\frac{\pi}{20}$,
$m_0=2000$ GeV and $m_{1/2}=500$ GeV.\\

\noindent
Fig.~(3a): Plot of the magnitudes of the electric dipole contribution,
of the color dipole contribution, of the purely gluonic contribution,
and of the total neutron EDM as a function of $|A_0|$ for the case when  
$\tb=3$, $\theta_{\mu0}= \frac{\pi}{30}$, $\alpha_{A0}=
\frac{\pi}{8}$, $m_{\tilde{g}}=800$ GeV ($m_{1/2}=281.2$ GeV)
and $m_0=1500$ GeV.\\

\noindent
Fig.~(3b): Same as Fig.(3a) except that 
$\alpha_{A0} = - \frac{\pi}{8}$.\\

\noindent
Fig.~(3c): Plot of the magnitudes of the neutron, the electron and the muon
EDMs  as a function of $m_{1/2}$ for the case when 
$\tb=3$, $\theta_{\mu0}= \frac{\pi}{30}$, $\alpha_{A0}=
\pm\frac{\pi}{8}$, $m_0=800$ Gev and $|A_0|=$ 2.6. The curve 1 (dotted) is 
for the case when $\alpha_{A0}=\frac{\pi}{8}$
and curve 2 (solid) is for case
 when $\alpha_{A0}=-\frac{\pi}{8}$.\\ 

\noindent 
Fig.~(3d): Plot of the magnitudes of the neutron EDM as a function of
$m_0$ for three cases 
 when $\theta_{\mu0}= \frac{\pi}{20}$,  
$\alpha_{A0}= -\frac{\pi}{6}$, and tan$\beta=3$. The data for the other
SUSY parameters is as follows: $|A_0|$=2.5, $m_{\tilde g}$=500 GeV for curve 1,
$|A_0|$=2.0, $m_{\tilde g}$=500 GeV for curve 2, and 
$|A_0|$=2.5, $m_{\tilde g}$=600 GeV for curve 3.\\

\noindent
Fig.~(3e): The excluded regions in the $m_0-m_{1/2}$ plane of the minimal
SUGRA model under the experimental constraints of Eqs.(44) and (45) when
$|A_0|=$ 1.4,  $\tb=3.0$,  $\theta_{\mu0}= \frac{\pi}{30}$ and
 $\alpha_{A0}=\pm\frac{\pi}{8}$.
The neutron
EDM curves are solid with $n({\pm})$ corresponding to 
$\alpha_{A0}=\pm\frac{\pi}{8}$,
and the electron EDM curve is dotted and labelled $e(+,-)$. 
The excluded regions
of the parameter space lie between the axes and the curves.

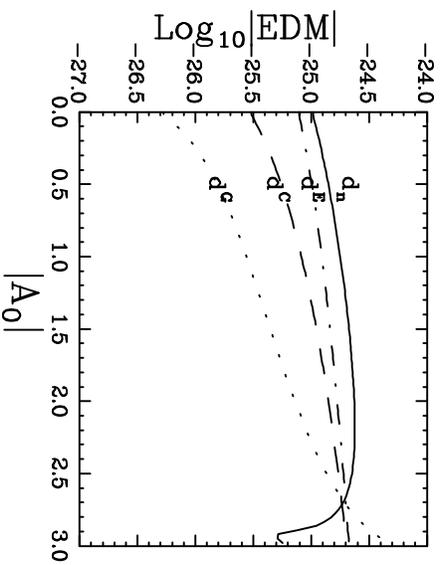
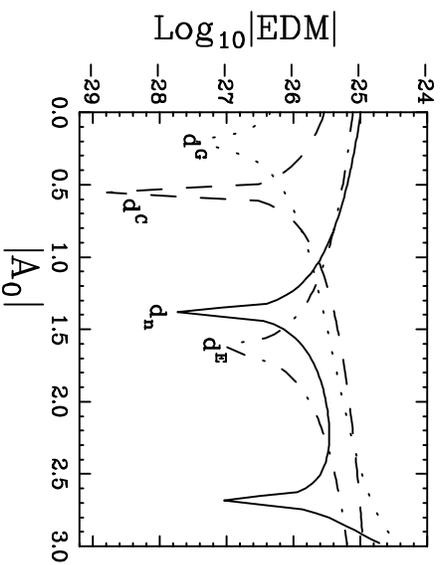
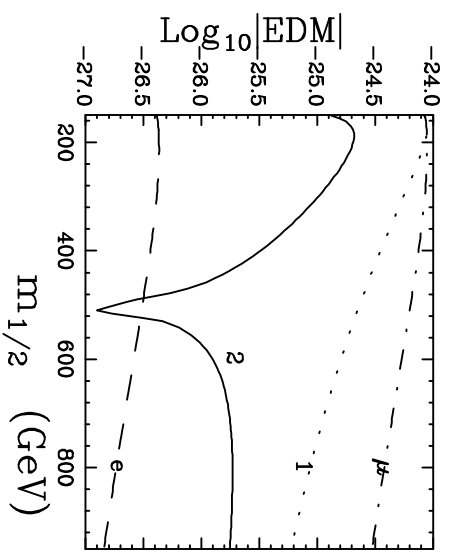
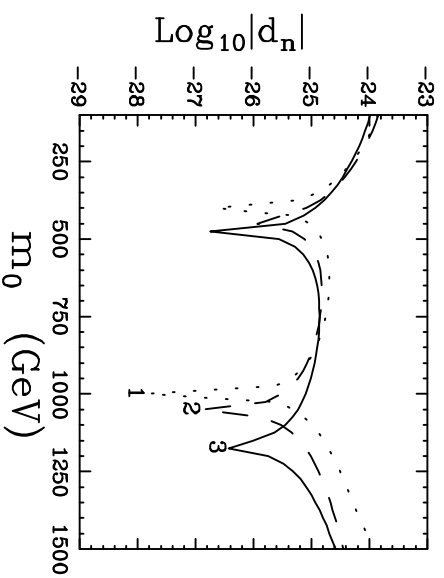
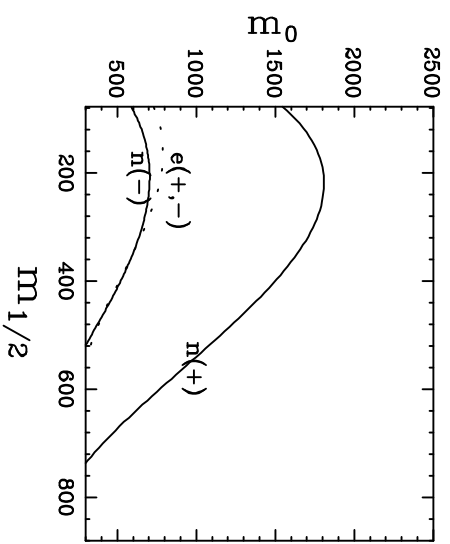
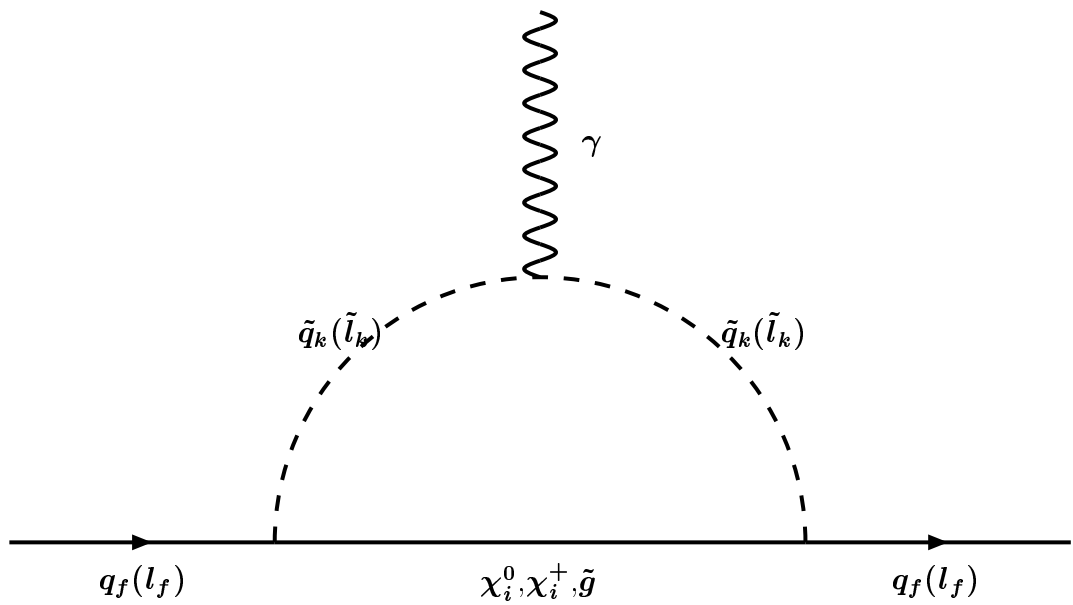
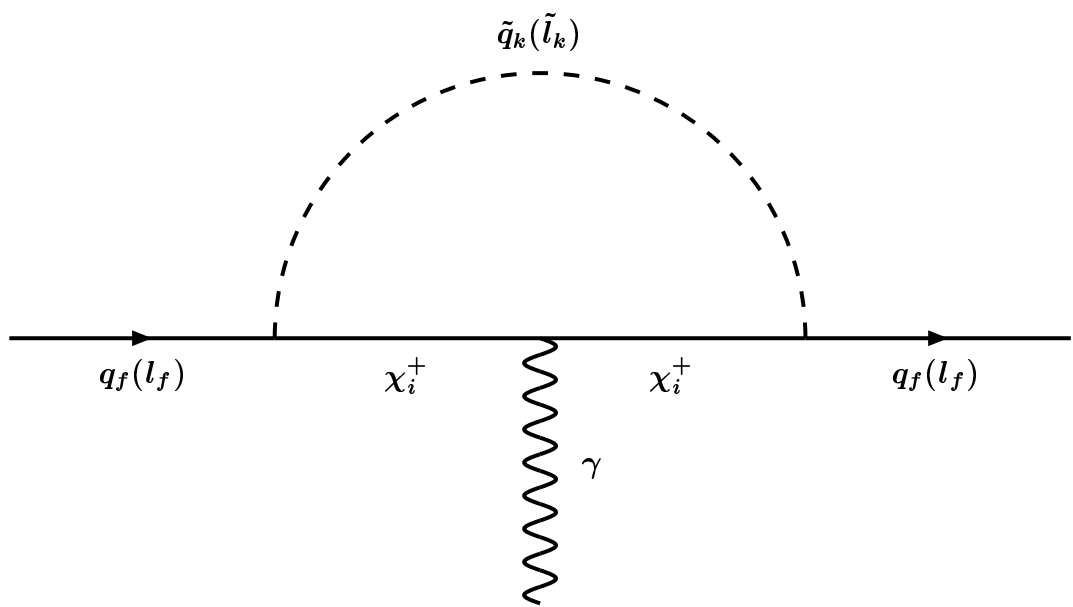
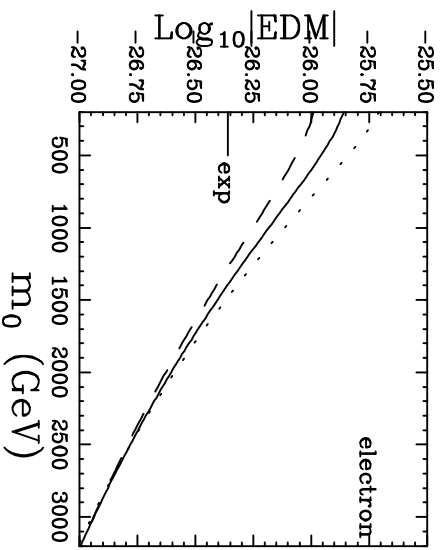
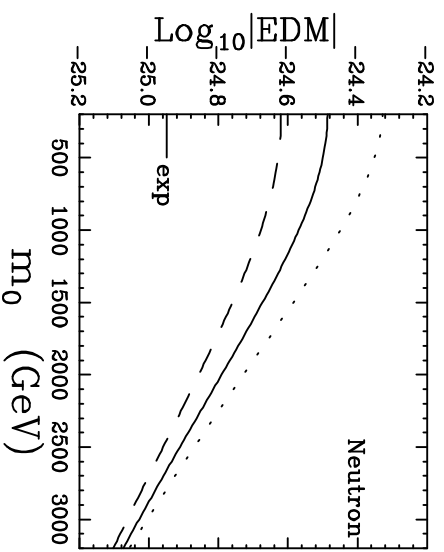
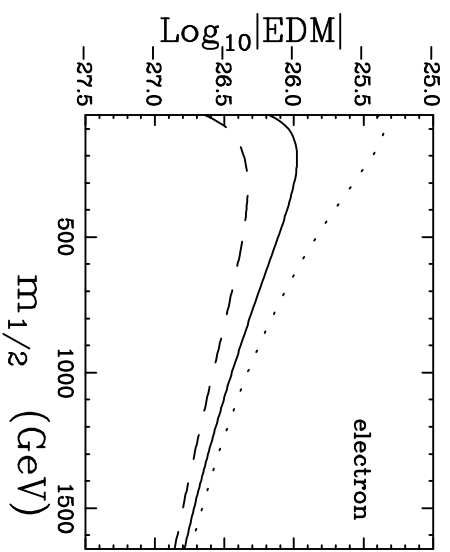
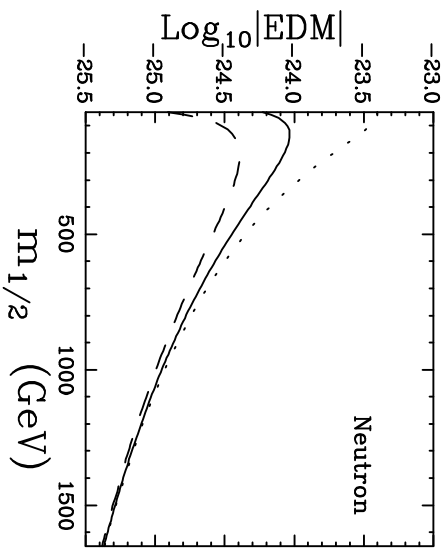
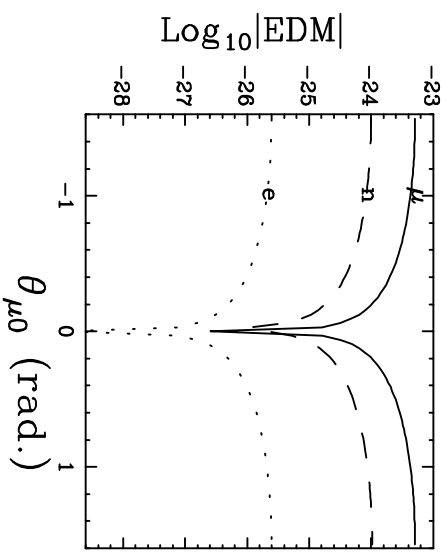
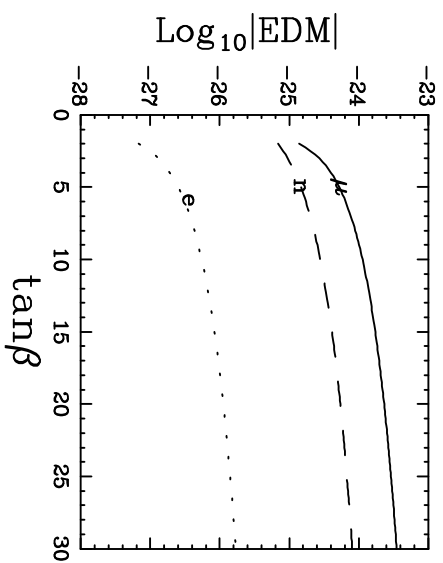
\end{document}